\documentclass[11pt,a4paper]{article}

\usepackage[T1]{fontenc}
\usepackage[utf8]{inputenc} % Remove if using LuaLaTeX or XeLaTeX
\usepackage{graphicx}
\usepackage{amsmath,amssymb}
\usepackage{cite}
\usepackage[colorlinks=true,linkcolor=blue,citecolor=blue,urlcolor=blue]{hyperref}
\usepackage{QCDEX2026}

\begin{document}

% Small workshop logo in the upper-left corner of the first page.
\noindent\includegraphics[width=0.4\textwidth]{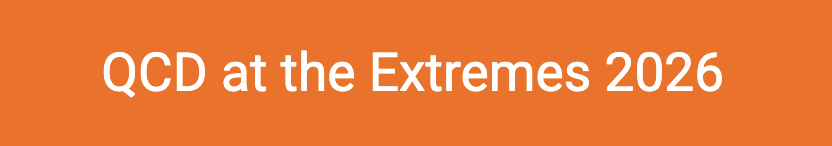}\par
\vspace{0.6em}
\begin{center}
  {\LARGE\bfseries 
 Parton branching TMD calculations 
with next-to-next-to-leading-logarithmic accuracy\par}  
  \vspace{0.8em}
  {\large
    A.~ Ben Mahmoud$^{1}$, 
    F.~Hautmann$^{2}$   
       and 
    A.~Lelek$^{1,*}$\\}
  \vspace{0.5em}
  {\small
    $^{1}$Universiteit Antwerpen, Antwerpen, Belgium\par
    $^{2}$University of Oxford, Oxford, United Kingdom\par}
  \vspace{0.35em}
  {\small $^{*}$Speaker \& Corresponding author: \href{mailto:aleksandra.lelek@uantwerpen.be}{aleksandra.lelek@uantwerpen.be}\par}
  \vspace{0.35em}
%  {\small\itshape refereed by the conveners of QCDEX2026\par}
\end{center}
\vspace{0.5em}

\begin{abstract}
We give  a concise description of recent  advances in the application of 
the parton branching approach to the QCD evolution of 
transverse momentum dependent 
(TMD) distributions,  particularly regarding  
the inclusion of next-to-next-to-leading-logarithmic contributions 
to Sudakov form factors. 
\end{abstract}

\noindent\textbf{Keywords:} QCD; collider physics; parton showers; parton distribution functions; Sudakov resummation  

\section{Introduction}

The quest for ever increasing precision  
at present and future particle colliders~\cite{Azzi:2019yne,LHeC:2020van,FCC:2018byv,Proceedings:2020eah} poses 
high demands on the performance of the 
parton-shower Monte 
Carlo generators~\cite{Buckley:2019kjt}
used for event simulations in physics analyses. 
Thus, a large effort is being   devoted to improving the 
logarithmic 
 accuracy of the parton 
 showers~\cite{Bewick:2019rbu,Nagy:2020rmk,Nagy:2020dvz,Forshaw:2020wrq,Holguin:2020joq,FerrarioRavasio:2023kyg,vanBeekveld:2024wws,vanBeekveld:2024qxs,Herren:2022jej,Gellersen:2021eci}  as well as to investigating the   
systematic uncertainties associated with 
parton shower computations, e.g.~from the treatment 
of parton distribution functions (PDFs) 
in Monte Carlo generators~\cite{Nagy:2020gjv,Collins:2002ey,Collins:2000gd,Hautmann:2007uw,prestel:2020,Collins:2005uv,Mendizabal:2023mel,Frixione:2023ssx,Hautmann:2025fkw,Jung:2025mtd} and from the kinematic reshuffling 
transformations to deal with 
the transverse momentum 
recoils~\cite{Bierlich:2022pfr,CASCADE:2021bxe,Sherpa:2019gpd,Bellm:2015jjp,Hautmann:2012dw,Dooling:2012uw,CASCADE:2010clj}.  

Refs.~\cite{Hautmann:2017fcj,Hautmann:2017xtx} 
propose the use of 
transverse momentum dependent 
(TMD) parton 
distributions~\cite{Angeles-Martinez:2015sea,Abdulov:2021ivr}  
in parton branching (PB) algorithms. This   
provides a complementary approach to the 
 framework~\cite{Aybat:2011zv,Collins:2011zzd,Collins:1984kg}  for 
TMD evolution, and  
  gives a method to incorporate effects of TMD 
dynamics in shower Monte Carlo generators, 
 designed to treat consistently both 
collinear PDFs and TMD distributions. 

\begin{figure}[htbp]
  \centering
  % Replace example-figure.pdf with your figure file and uncomment the next line.
   \includegraphics[width=0.52\textwidth]{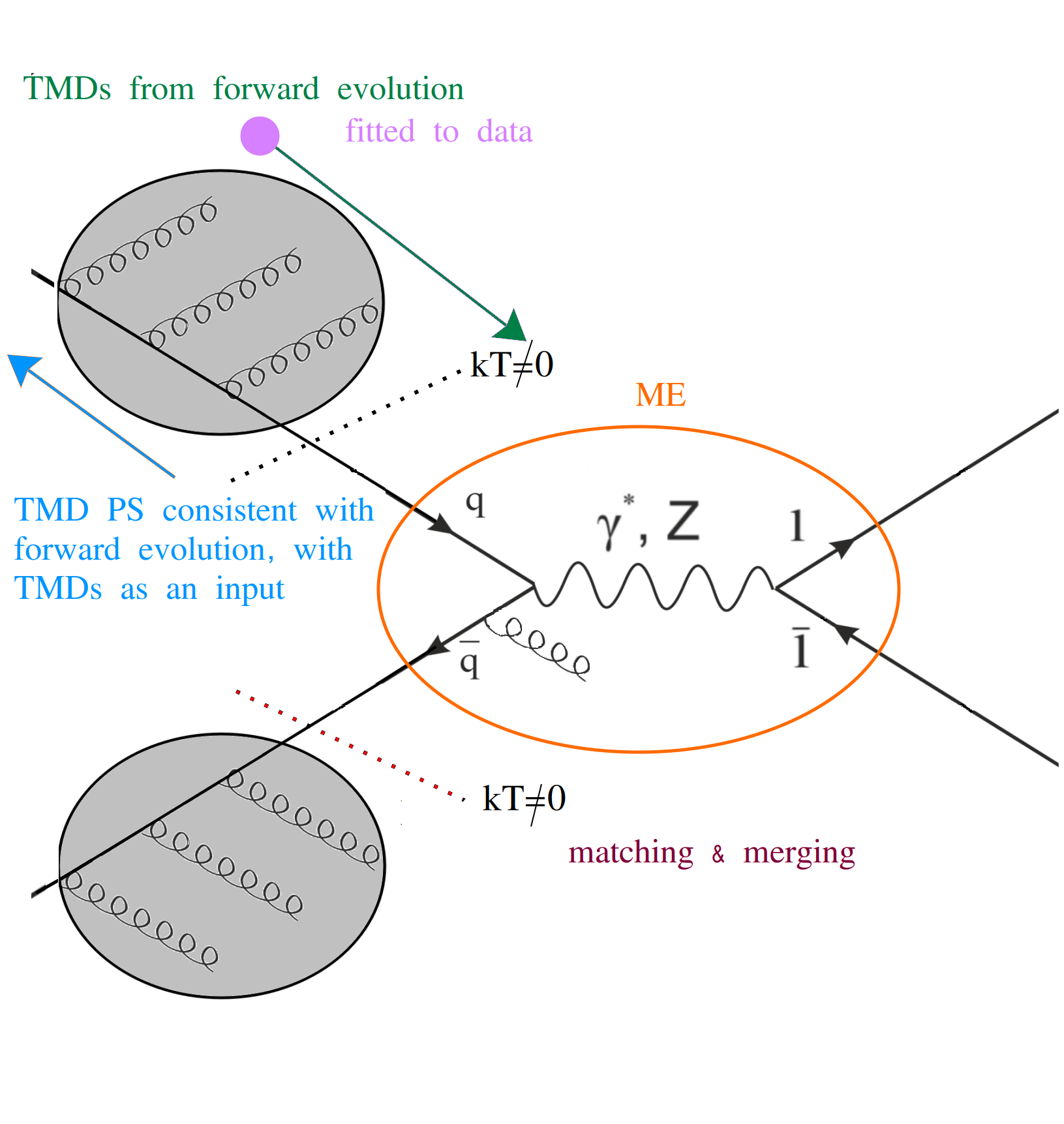}
%  \fbox{\parbox[c][35mm][c]{0.72\textwidth}{\centering Replace this box with your figure.}}
  \caption{PB TMD picture of Drell-Yan lepton-pair production.}
  \label{fig:fig1}
\end{figure}

Fig.~\ref{fig:fig1} pictures schematically the PB TMD approach 
for the case of the 
Drell-Yan (DY) lepton-pair hadro-production process. 
The hard scattering matrix element on the right hand side, 
evaluated at finite order of 
perturbation theory~\cite{Alwall:2014hca}, is combined, through 
matching and merging techniques~\cite{BermudezMartinez:2019anj,Yang:2022qgk,BermudezMartinez:2021lxz,BermudezMartinez:2022bpj}, with the  TMD 
parton distributions and showers~\cite{BermudezMartinez:2018fsv,CASCADE:2021bxe} undergoing 
QCD evolution on the left hand side. 

 In Fig.~\ref{fig:fig1}, 
both perturbative and non-perturbative contributions to 
the TMD distributions are to be taken into account. 
Perturbatively, the TMD evolution is controlled by 
splitting functions and Sudakov form factors.   These are 
evaluated through power series expansions in the strong coupling 
$\alpha_s$~\cite{Hautmann:2017fcj}, and achieve 
the all-order resummation of perturbative logarithmically-enhanced 
corrections in the Sudakov  region of 
small transverse momenta~\cite{Collins:1984kg} 
(besides the resummation of collinear 
DGLAP~\cite{Gribov:1972ri,Altarelli:1977zs,Dokshitzer:1977sg} 
logarithms).  
Nonperturbative contributions, on the other hand, 
 involve both 
the TMD distributions, for all flavors, at the low scale 
of the branching evolution, of order 1 GeV,  and 
the large-distance components of the Sudakov 
form factors~\cite{Bubanja:2023nrd,Hautmann:2025fkw,Favart:2026wfc}.

Most PB TMD applications have been worked through so far at 
leading-logarithmic (LL) and next-to-leading-logarithmic (NLL) order in the Sudakov resummation  
(with leading order (LO) and  next-to-leading order (NLO) in the collinear DGLAP resummation).  
The extension to next-to-next-to-leading logarithmic (NNLL) order has 
been treated recently in~\cite{Martinez:2024mou,ALelekEtAll}. 
In what follows we give a brief account of current NNLL results.

\section{Soft-gluon coupling and TMD distributions}

The PB formulation of the evolution of TMD parton distributions has been 
successful in incorporating effects of TMD dynamics in Monte Carlo 
predictions for collider observables ranging from the 
vector boson $p_T$ 
spectrum~\cite{BermudezMartinez:2019anj,Bubanja:2023nrd,BermudezMartinez:2020tys,Zhan:2024lym,Moureaux:2025cyi,Favart:2026wfc,Favart:2026vcn} 
to multi-jet cross 
sections~\cite{BermudezMartinez:2021lxz,BermudezMartinez:2022bpj,BermudezMartinez:2021zlg,Abdulhamid:2021xtt,Yang:2022qgk,BermudezMartinez:2020qta} 
to deep inelastic scattering  (DIS)
structure functions~\cite{BermudezMartinez:2018fsv,Hautmann:2025fkw}, 
and has been widely applied with NLL accuracy in the Sudakov region. 

\begin{figure}[htbp]
  \centering
  % Replace example-figure.pdf with your figure file and uncomment the next line.
   \includegraphics[width=0.5\textwidth]{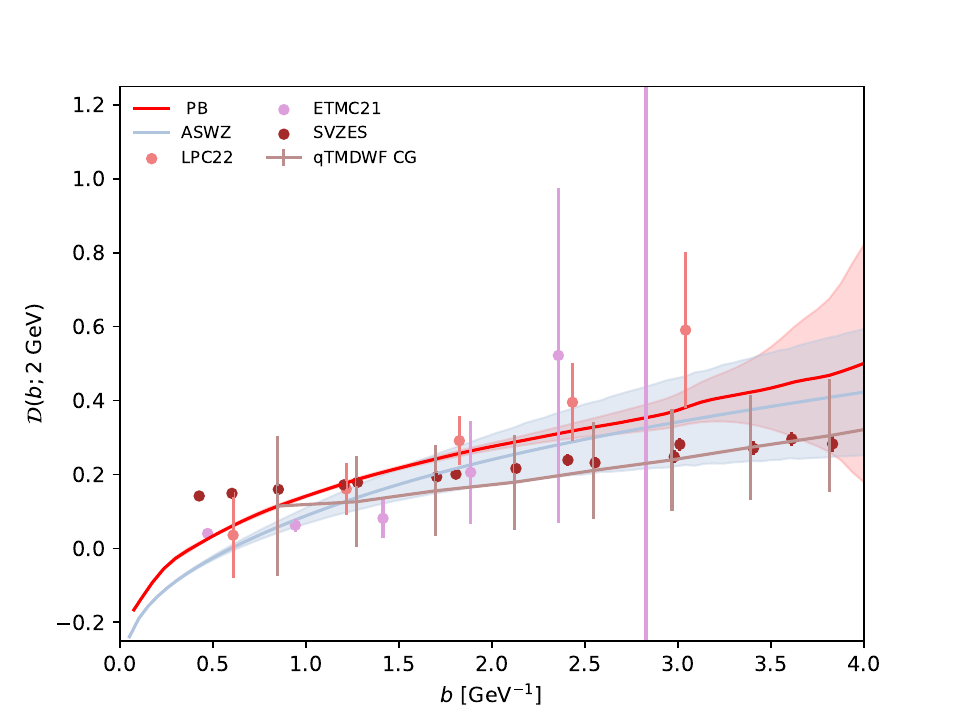}
%  \fbox{\parbox[c][35mm][c]{0.72\textwidth}{\centering Replace this box with your figure.}}
  \caption{PB TMD result for the rapidity evolution kernel 
  as a function of the transverse distance $b$, compared with 
  results from lattice 
  computations~\protect\cite{LPC:2022ibr,Schlemmer:2021aij,Li:2021wvl,Avkhadiev:2024mgd,Avkhadiev:2023poz,Bollweg:2024zet}.}
  \label{fig:fig7}
\end{figure}

The PB TMD method implies evolution in 
both mass and rapidity, in which the 
soft-gluon resolution scale~\cite{Hautmann:2017xtx} 
plays the role of a rapidity cut-off.  
The rapidity evolution kernel~\cite{Collins:1981va,Collins:1981uk}  
can be computed 
by  the technique used in~\cite{Martinez:2024mou,Martinez:2024twn}.  
Results are shown in Fig.~\ref{fig:fig7} 
for the dependence of this kernel on the transverse coordinate $b$, 
conjugate to transverse momentum, at fixed value of the evolution 
scale $\mu$. For comparison, in Fig.~\ref{fig:fig7}       
results are also shown from lattice-QCD  
computations~\cite{LPC:2022ibr,Schlemmer:2021aij,Li:2021wvl,Avkhadiev:2024mgd,Avkhadiev:2023poz,Bollweg:2024zet} in the 
literature.  A  substantial agreement across 
the $b$ range, within the large theoretical 
uncertainties, is observed between the different 
calculations.

One of the difficulties in extending this approach to include 
NNLL contributions has been the mismatch, which starts at NNLL, 
between the double-logarithmic coefficient of the Sudakov form factor 
and the residue of the collinear splitting functions
at the soft-gluon pole~\cite{Becher:2010tm}.    
  In Refs.~\cite{Martinez:2024mou,Martinez:2024twn,Lelek:2025enb} this 
  difficulty has been 
  overcome thanks to the use of  the so-called 
  ``soft-gluon physical 
  coupling''~\cite{Banfi:2018mcq,Catani:2019rvy} in 
  the parton branching algorithm.

The key idea to achieve  NNLL accuracy is to 
use the  
soft-gluon physical coupling~\cite{Banfi:2018mcq,Catani:2019rvy,Catani:1990rr} 
via  the transformation 
$ \alpha_s \to    \alpha_s^{\text{phys}}$, with 
\begin{equation} \label{eq:effectivecoupling}
    \alpha_s^{\text{phys}}=\alpha_s \left( 1 + \sum_{n=1}^\infty \mathcal{K}^{(n)} \left( \frac{\alpha_s}{2\pi}\right)^n  \right)\;, 
\end{equation}
where the first-order coefficient  is given in~\cite{Catani:1990rr}, and 
 the  second-order coefficient is given in~\cite{Banfi:2018mcq,Catani:2019rvy}.  
In order to avoid the double counting,  
at order $\alpha_s^2$,  between  the coefficient $\mathcal{K}^{(1)}$ in Eq.~(\ref{eq:effectivecoupling}) 
and the cusp contribution in the NLO splitting function, 
we use the  subtracted form of the physical coupling 
$\alpha_s^{\text{subtr}} = 
\alpha_s^{\text{phys}} - \mathcal{K}^{(1)} \alpha_s^2 / ( 2\pi)$. 
The product of $\mathcal{K}^{(2)}$ with the cusp part of the LO splitting function 
$k_a^{(0)}$
provides the missing NNLL double-logarithmic  
coefficient, i.e. $\mathcal{K}^{(2)} \cdot k_a^{(0)} = A_a^{(3)}$. 
As regards the NNLL single-logarithmic coefficient  $B^{(2)}$, on the 
other hand, this is supplied by the two-loop 
splitting functions in the PB TMD evolution,
and is affected by the scheme dependence
arising from the renormalization group equation, as 
described in~\cite{Lelek:2025enb}.

The details of the PB implementation can be found in \cite{Martinez:2024mou,Lelek:2025enb,ALelekEtAll,BenMahmoud:2026lvl}. In the calculations, we investigate two prescriptions for the soft-gluon resolution scale $z_M$, i.e., 
fixed~\cite{BermudezMartinez:2018fsv} and 
dynamical~\cite{Hautmann:2019biw} $z_M$. 
The calculations are performed with the scale of $\alpha_s$ set 
equal to the emitted transverse momentum $q_{\bot}$.
We study  two values of the parameter $q_s$ defining the intrinsic-$k_{\bot}$ Gaussian width, $q_s=0.5$ and $q_s=1.0$ GeV, which controls the TMD distribution for 
low mass scales in the QCD evolution but does not affect the integrated 
TMD distributions (iTMD).

\begin{figure}[htbp]
  \centering
  % Replace example-figure.pdf with your figure file and uncomment the next line.
   \includegraphics[width=0.6\textwidth]{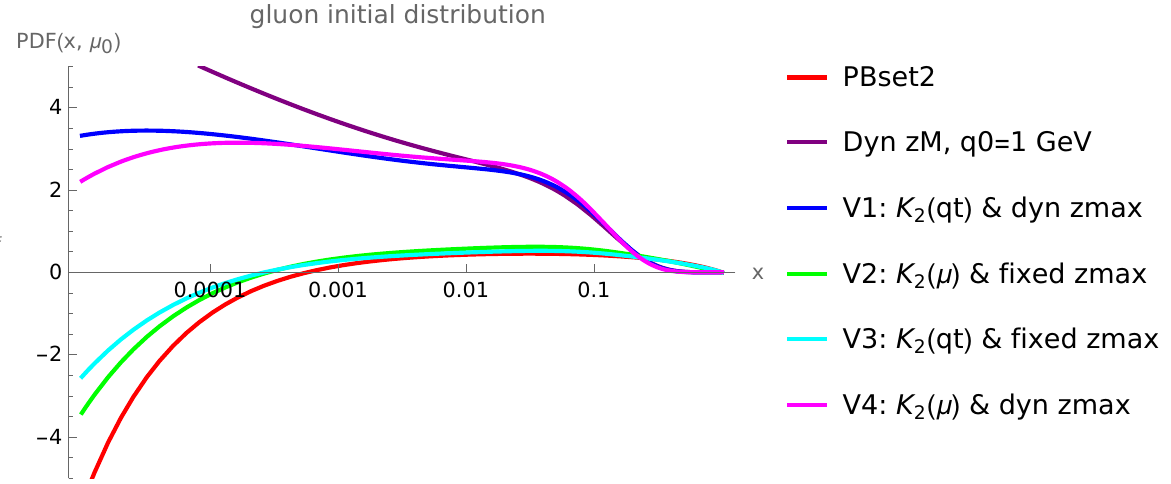}
   \includegraphics[width=0.37\textwidth]{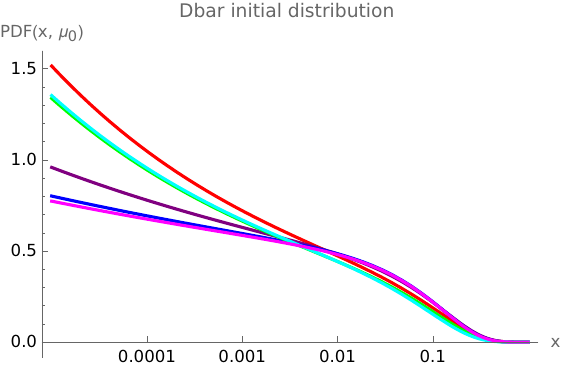}
   \includegraphics[width=0.37\textwidth]{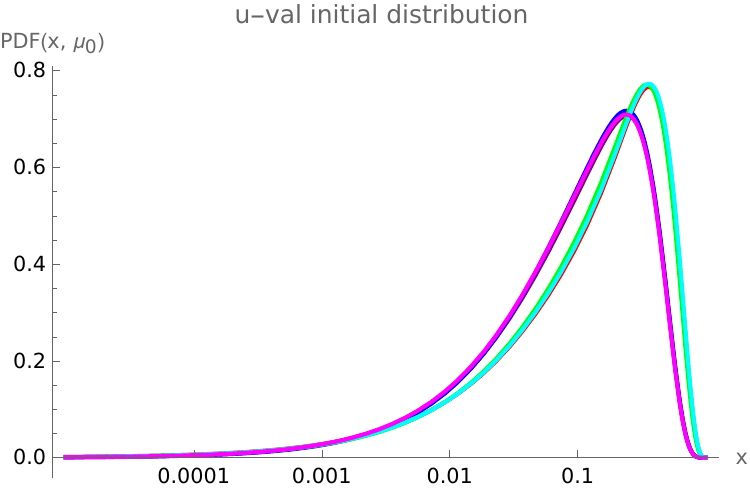}
%  \fbox{\parbox[c][35mm][c]{0.72\textwidth}{\centering Replace this box with your figure.}}
  \caption{Gluon, ${\bar d}$ and valence $u$-quark distributions from NNLL TMD 
  determinations based on precision DIS fits.}
  \label{fig:fig3}
\end{figure}

\subsection{Number of active quark flavors}   

It is worth noting that 
the scale choice in $\alpha_s$ introduces an additional ambiguity in the treatment of the number of active quark flavours $n_f$, i.e. the threshold for $n_f\rightarrow n_f+1$. 

The default PB setup to treat number of active quark flavors ($n_f$) is the following.
In the splitting functions $P_{ab}$, which explicitly depend on $n_f$, the mass threshold is calculated based on the branching scale $\mu^{\prime}$, i.e., if $\mu^{\prime} > m_i$, where $m_i$ is the mass of quark of flavour $i$,  the quark of flavor $i$ can be produced in the splitting process and it is accounted for in the number of active flavors. 
However, there is also an implicit dependence on $n_f$ via $\alpha_s$. If the transverse momentum $q_{\bot}$ is used as a scale in $\alpha_s$, the number of active flavours in the  $\alpha_s$ evaluation is calculated based on $q_{\bot}$. 
To illustrate this, the splitting function can be written in a more explicit form
\begin{equation}
    P_{ab}(z, \mu^{\prime}) = \sum_{n=1}^{\infty} \left(\frac{\alpha_s(n_f(q_{\bot}))}{2\pi}\right)^{n}P^{(n-1)}_{ab}(z, n_f(\mu^{\prime}))\;.
\end{equation}
The $n_f$ treatment influences the use of the soft-gluon coupling 
 $\alpha_s^{\rm{phys}}$: the coefficients $\mathcal{K}^{(n)}$ depend 
 on $n_f$ and one can calculate these based on the 
 emitted transverse momentum $q_{\bot}$, i.e., 
$  \alpha_s^{\text{phys}}(q_{\bot})=\alpha_s(q_{\bot}) \left( 1 + \sum_{n=1}^\infty \mathcal{K}^{(n)}(q_{\bot}) \left( {\alpha_s(q_{\bot})} / {(2\pi)}\right)^n  \right)$, 
or on the branching scale $\mu^{\prime}$, i.e., 
$
    \alpha_s^{\text{phys}}(q_{\bot}, \mu^{\prime})=\alpha_s(q_{\bot}) \left( 1 + \sum_{n=1}^\infty \mathcal{K}^{(n)} (\mu^{\prime})\left( {\alpha_s(q_{\bot})} / {(2\pi)}\right)^n  \right) . 
$
We thus investigate the impact of the new NNLL term 
on the parton distributions and DY predictions 
 in two scenarios, i.e., with $\mathcal{K}^{(2)}(n_f(q_{\bot}))$ or with 
  $\mathcal{K}^{(2)}(n_f(\mu^{\prime}))$.

  \subsection{Fit results}

Using PB TMD evolution at NNLL, Refs.~\cite{BenMahmoud:2026lvl,ALelekEtAll} perform
determinations of iTMD distributions based on xFitter \cite{xFitter:2022zjb,Alekhin:2014irh} fits to
the precision measurements~\cite{H1:2015ubc} of
DIS structure functions, in a similar
framework to the
NLO determinations of
Refs.~\cite{BermudezMartinez:2018fsv,Hautmann:2025fkw}.
Results for the gluon, ${\bar d}$   and $u$  quark  distributions
are shown in Fig.~\ref{fig:fig3} as functions of $x$, at the starting 
evolution scale, for different configurations
of the soft-gluon coupling and resolution scale as described above, 
and compared with the reference 
NLO PB TMD sets \cite{BermudezMartinez:2018fsv,Hautmann:2025fkw}.
We note the difference between the fixed and 
dynamical $z_M$ cases, and the NNLL effects in both cases.

\begin{figure}[htbp]
  \centering
  % Replace example-figure.pdf with your figure file and uncomment the next line.
     \includegraphics[width=0.35\textwidth]{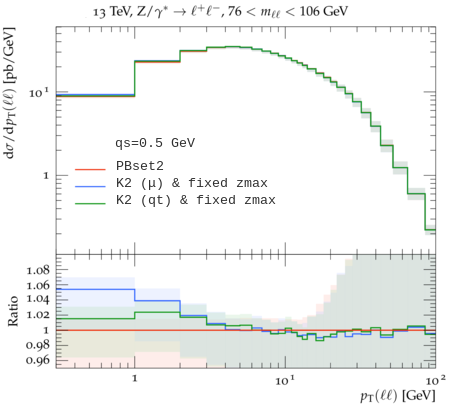}
   \includegraphics[width=0.35\textwidth]{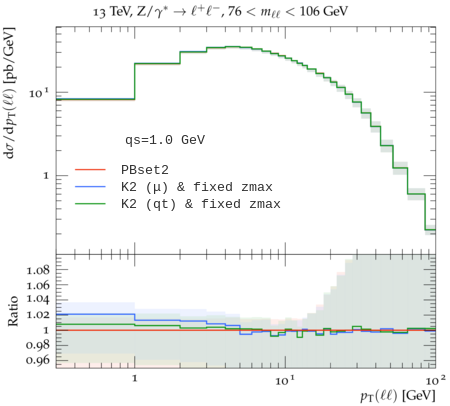}
%  \fbox{\parbox[c][35mm][c]{0.72\textwidth}{\centering Replace this box with your figure.}}
  \caption{Influence of the NNLL contributions on the DY transverse momentum spectrum.}
  \label{fig:fig6}
\end{figure}

\section{Drell-Yan lepton-pair production}

Transverse momentum spectra 
of vector bosons  via DY lepton-pair decay 
are  sensitive to TMD parton distributions.  
Experimental measurements of DY transverse momentum $p_T$,   
at the LHC as well as  at 
lower-energy experiments, are used for TMD extractions  
in 
Refs.~\cite{Moos:2025sal,Barry:2025glq,Bacchetta:2024qre,Bury:2022czx},  
employing the approach~\cite{Aybat:2011zv,Collins:2011zzd,Collins:1984kg} 
to TMD evolution, 
and in  Refs.~\cite{Bubanja:2023nrd,Favart:2026vcn},  
employing  the PB TMD approach~\cite{Hautmann:2017xtx,Hautmann:2017fcj}. 
They are also  used in Ref.~\cite{CMS:2024goo} 
for the determination of 
the intrinsic transverse momentum 
$k_T$   
from the tuning of parton-shower 
Monte Carlo event 
generators~\cite{Bierlich:2022pfr,Bellm:2015jjp} to experimental data.  

In Fig.~\ref{fig:fig6} we examine the impact of 
NNLL contributions on PB TMD predictions for the $Z$-boson 
transverse momentum distribution at LHC energies. 
In each plot, the NLO curve (PBset2) is compared with 
two NNLL predictions, with two different $n_f$ treatments.

We see a visible impact of NNLL contributions in small $p_{\bot}$ region,
with a significant role played by the $n_f$ treatment. 
 Increasing the Gaussian width reduces these effects, 
 as illustrated by the comparison of the left and right panels in Fig.~\ref{fig:fig6}.
The effects also become less pronounced for dynamical $z_M$ scenarios.

\section{Conclusion}

PB TMD calculations have been successfully applied to the 
description of electroweak boson production spectra 
and jet cross sections in hadronic collisions  as well as to 
the description of measurements of 
deep inelastic scattering structure functions.  
In this article we have discussed the extension 
of the PB TMD method to the 
NNLL order of  accuracy 
in the Sudakov region. 

In the case of double-logarithmic 
resummations (such as Sudakov resummations),   
it may be argued that LL and NLL orders 
serve to provide a reliable evaluation of 
the central values of the predictions, while 
the NNLL order enables one to reliably 
access the uncertainty band on the predictions. 
The advances described in this article thus 
pave the way to control the perturbative 
accuracy of the results, do precision physics both in 
the TMD quark and gluon sectors, and explore 
non-perturbative effects encoded in TMD 
parton distributions and in the rapidity evolution 
kernels.

\section*{Acknowledgements}
We thank A.~Bermudez Martinez, L.~Favart, 
H.~Jung, L.~Keersmaekers, L.~Moureaux, 
M.~Mendizabal Morentin, N.~Raicevic, S.~Sadeghi Barzani, 
S.~Taheri Monfared and  A.M.~van Kampen 
for collaboration 
and discussion. AL acknowledges funding by 
Research Foundation - Flanders (FWO) ( 1278325N ).

\bibliographystyle{QCDEX2026}
\bibliography{qcdex2026_bibl}

\end{document}